\documentclass{aa}
\usepackage{graphicx}
\usepackage{txfonts}

\begin{document}

\title{Molecular emission from GG\,Car's circumbinary disk\thanks{Based on 
observations collected with the ESO VLT Paranal Observatory under programs 
384.D-0613(A) and 088.D-0442(B)}}
%\subtitle{A pre-merger object or a single star with an eclipsing disk?}

\author{M. Kraus\inst{1}, M.E. Oksala\inst{1}, D.H. Nickeler\inst{1}, 
M.F. Muratore\inst{2,3}, M. Borges Fernandes\inst{4}, A. Aret\inst{5,1}, 
L.S. Cidale\inst{2,3},
\and
W.J. de Wit\inst{6}
  }
                                                                                
\institute{Astronomick\'y \'ustav, Akademie v\v{e}d \v{C}esk\'e republiky,
Fri\v{c}ova 298, 251\,65 Ond\v{r}ejov, Czech Republic\\
\email{kraus@sunstel.asu.cas.cz}
\and
Departamento de Espectroscop\'ia Estelar, Facultad de Ciencias
Astron\'omicas y Geof\'isicas, Universidad Nacional de La Plata, Paseo del 
Bosque s/n, B1900FWA, La Plata, Argentina
%\email{fmuratore@carina.fcaglp.unlp.edu.ar}
\and
Instituto de Astrof\'isica de La Plata, CCT La Plata, CONICET-UNLP,
Paseo del Bosque s/n, B1900FWA, La Plata, Argentina
\and
Observat\'orio Nacional, Rua General Jos\'e Cristino 77,
20921-400 S\~ao Cristov\~ao, Rio de Janeiro, Brazil
\and
Tartu Observatory, 61602, T\~oravere, Tartumaa, Estonia
\and
European Southern Observatory, Alonso de Cordova 3107, Vitacura, Santiago, 
Chile
      }
                                                                                
\date{Received; accepted}
                                                                                
\authorrunning{Kraus et al.}
%\titlerunning{Circumbinary disk of GG Car}

\abstract
%Context
{The appearance of the B[e] phenomenon in evolved massive stars such as
B[e] supergiants is still a mystery. While these stars are generally 
found to have disks that are cool and dense enough for efficient molecule
and dust condensation, the origin of the disk material is still unclear.}
%Aims
{We aim at studying the kinematics and origin of the disk in the eccentric 
binary system GG\,Car, whose primary component is proposed to be a B[e]
supergiant.}
%Methods
{Based on medium- and high-resolution near-infrared spectra we analyzed the 
CO-band emission detected from GG\,Car. The complete CO-band structure
delivers information on the density and temperature of the emitting region,
and the detectable \element[][13]{CO} bands allow us to constrain the 
evolutionary phase. In addition, the kinematics of the CO gas can be extracted 
from the shape of the first \element[][12]{CO} band head.} 
%Results
{We find that the CO gas is located in a ring surrounding the 
eccentric binary system, and its kinematics agrees with 
Keplerian rotation with a velocity, projected to the line of sight, of
$80\pm 1$\,km\,s$^{-1}$. The CO ring has a column density of $(5\pm 3)\times
10^{21}$\,cm$^{-2}$ and a temperature of $3200\pm 500$\,K.
% and, assuming it is co-planar to the orbital plane, an inclination of 
% $63\degr\pm 9\degr$. 
In addition, the material is chemically enriched in 
\element[][13]{C}, which agrees with the primary component being 
slightly evolved off the main sequence. We discuss two possible 
scenarios for the origin of the circumbinary disk: (i) non-conservative
Roche lobe overflow, and (ii) the possibility that the progenitor of the 
primary component could have been a classical Be star. Neither
can be firmly excluded, but for Roche lobe overflow to occur,
a combination of stellar and orbital parameter extrema 
would be required.} 
%Conclusions
{}

\keywords{Stars: emission-line, Be -- Stars: early-type -- supergiants -- 
circumstellar matter -- Stars: individual: \object{GG Car}}

\maketitle

\section{Introduction}

The Galactic emission line star GG\,Car was discovered more than a century ago 
(Pickering \cite{Pickering}; Pickering \& Fleming \cite{PickeringFleming}). 
In addition to strong Balmer line emission, its optical spectrum displays numerous 
emission lines from permitted and forbidden transitions of 
predominantly low-ionized metals (e.g. Swings \cite{Swings}; Marchiano et al. 
\cite{Paula}). Strong line profile variation of emission features in the 
optical (e.g. Machado et al. \cite{Dora}) and near-infrared (Morris et al. 
\cite{Morris}) was reported, which might be interpreted as variability
in the density structure of the wind. Furthermore, the spectral energy 
distribution shows a pronounced infrared excess, 
characteristic of emission from hot circumstellar dust (Allen \cite{Allen};
Cohen \& Barlow \cite{Cohen}; Bouchet \& Swings \cite{Bouchet}). These
reported characteristics, together with the temperature (Lopes et al. 
\cite{Lopes}) and luminosity estimates (McGregor et al. \cite{McGregor}),
resulted in the classification of GG\,Car as a B[e] supergiant (B[e]SG;
Lamers et al. \cite{Lamers}), although its supergiant status was long unclear 
due to highly uncertain distance estimates. 

GG\,Car displays radial velocity and light variability in both the
optical and UV light curves, suggestive of an eclipsing binary system
(Hern\'{a}ndez et al. \cite{Hernandez}; Gosset et al. \cite{Gosset84,Gosset85};
Brandi et al. \cite{Brandi}). Based on radial velocity measurements of the
blueshifted absorptions of the \ion{He}{i} P\,Cygni profiles, Marchiano et al.
(\cite{Paula}) recently refined the period of this eccentric binary system
to $31.033\pm 0.008$\,d. In addition, these authors obtained orbital parameters
for the two components, as well as improved
stellar parameters for the primary, summarized in Table\,\ref{Tab:param}.
The mass of the primary and its temperature estimate confirm 
the supergiant status of GG\,Car. With this knowledge, the analysis of 
Marchiano et al. (\cite{Paula}) also provided the distance
of $5\pm 1$\,kpc toward GG\,Car based on two independent methods. This
distance estimate places GG\,Car 
in the vicinity of other evolved massive stars such as the luminous blue
variables (LBVs) HR\,Car ($d=5\pm 1$\,kpc; van Genderen et al.
\cite{vanGenderen}; Hutsem\'{e}kers \& van Drom \cite{Hutsemekers}) and AG\,Car
($d=6\pm 1$\,kpc; Humphreys et al. \cite{Humphreys}), but keeping in mind
that distance estimates generally have high uncertainties.

\begin{table}
\caption{Stellar parameters of the primary component of GG\,Car from
Marchiano et al. (\cite{Paula}).}
\label{Tab:param}
\centering
  \begin{tabular}{l l}
    \hline
    \hline
 $T_{\rm eff}$ [K]      & $23\,000\pm 2\,000$ \\
 $L/L_{\odot}$          & $(2.6\pm 1.0)\times 10^{5}$ \\
 $R_{*}$ [$R_{\odot}$]  & $32\pm 8$ \\
 $M$ [$M_{\odot}$]      & $26\pm 4$ \\
 $i$ [$\degr$]          & $63\pm 9$ \\
 $d$ [kpc]              & $5\pm 1$  \\
 $E(B-V)$ [mag]         & $0.51\pm 0.15$ \\
% $ $ [K]       &   [$\times 10^{5}$]    & [$R_{\odot}$] & [$M_{\odot}$]
% & [$\degr$] & [kpc]   & [mag] \\
%   \hline
% $23\,000\pm 2\,000$  & $2.6\pm 1.0$ &  $32\pm 8$ & $26\pm 4$ & $63\pm 9$ & $5\pm 1$ & $0.51\pm 0.15$ \\
% $\pm 2\,000$ & $\pm 1.0$     & $\pm 8$       & $\pm 4$ & $\pm 9$   & $\pm 1$ & $\pm 0.15$ \\
    \hline
  \end{tabular}
\end{table}

\begin{figure*}[t!]
\resizebox{\hsize}{!}{\includegraphics{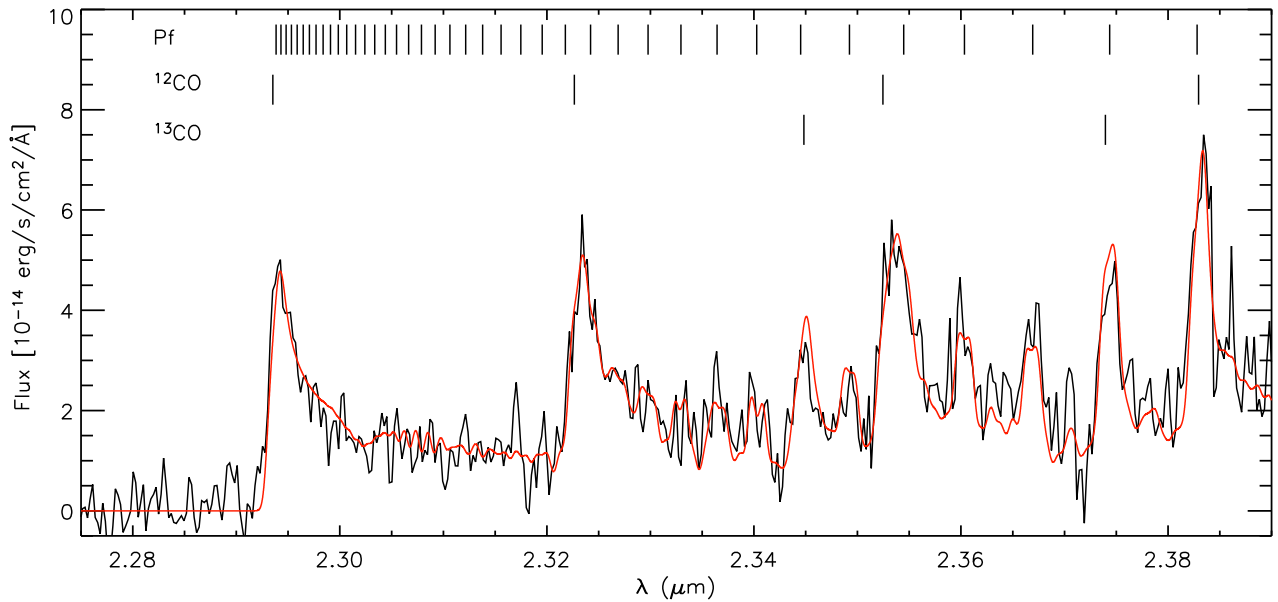}}
\caption{Fit (red/gray) to the flux-calibrated, continuum-subtracted SINFONI
spectrum (black) of GG\,Car. Positions of the CO band heads and hydrogen Pfund
lines are indicated.}
\label{Fig:SINFONI}
\end{figure*}

Optical polarimetric (Gnedin et al. \cite{Gnedin}; Klare \& Neckel 
\cite{Klare}; Barbier \& Swings \cite{Barbier}) as well as interferometric 
(Borges Fernandes \cite{Marcelo}) observations confirmed that the dust 
distribution is non-shperically symmetric, most probably located within a 
disk or disk-like structure. Preliminary analysis of the interferometric data 
deliver an inclination angle of about $60\degr$, which excellently 
agrees with the inclination of the orbital plane of $54\degr - 72\degr$ 
found by Marchiano et al. (\cite{Paula}). Furthermore, H$\alpha$ 
spectropolarimetric studies of Pereyra et al. (\cite{Pereyra}) revealed 
a rotating disk-like structure. 
A disk is an ideal environment, not only for 
dust, but also for molecules, such as the very robust CO molecule. Emission in 
CO bands from GG\,Car has been reported by McGregor et al. (\cite{McGregor}) 
and Morris et al. (\cite{Morris}). 

Kraus (\cite{Kraus09}) reported on the detection of \element[][13]{CO}-band 
emission from GG\,Car based on a low-resolution ($R\simeq 1500$) spectrum 
obtained with AMBER at the VLTI. The strength of the isotopic bands indicated 
that the circumstellar material is chemically enriched with processed material 
that must have been transported to the stellar surface and released into the 
environment by either strong non-spherically symmetric stellar winds or mass 
ejection events. The observed enhancement with \element[][13]{C}, however, 
strictly disagrees with theoretical model predictions of both rotating and 
non-rotating stellar models, and the disagreement could only be solved by the 
postulation of a higher mass (and thus luminosity) of GG\,Car than those 
reported in the literature. The new stellar parameters determined by Marchiano 
et al. (\cite{Paula}; see Table\,\ref{Tab:param}) moved the position of GG\,Car 
in the Hertzsprung-Russell (HR) diagram to a more appropriate place. Still, the 
proper determination of the amount of \element[][13]{CO} and accordingly the 
\element[][12]{C}/\element[][13]{C} ratio, which is an excellent tracer of the 
stellar evolutionary phase (Kraus \cite{Kraus09}; Liermann et al. 
\cite{Liermann}), is lacking. 

Probing the evolutionary stage of B[e]SGs is an important issue for 
understanding their mass-loss history, and hence the origin and shaping 
mechanism of their circumstellar environment resulting in the B[e] phenomenon.
For this, the amount of chemically enriched matter is essential, as well as
the geometry and kinematics of the circumstellar material. The goal of 
the present paper is, therefore, to extract these based on a detailed study of 
CO-band emission in medium- and high-resolution $K$-band spectra of the 
Galactic B[e]SG star GG\,Car.

\section{Observation and data reduction}\label{obs}

A medium-resolution ($R = 4500$) $K$-band ($1.95-2.45\,\mu$m) spectrum 
for the Galactic B[e] star GG\,Car was obtained on 2012 January 9, using the 
Spectrograph for INtegral Field Observation in the Near-Infrared (SINFONI;
Eisenhauer et al. \cite{Eisenhauer}; Bonnet et al. \cite{Bonnet}) on the
VLT UT4 telescope. The observations were performed with an AB nod pattern and
an $8\times 8$\,arcsec$^{2}$ field of view. For telluric correction and flux
calibration, a B-type standard star was observed at similar airmass.
The data were reduced with the SINFONI pipeline (version 2.2.9).
 
For flux calibration, we applied a Kurucz model (Kurucz \cite{Kurucz}) to the
standard star spectrum and scaled it to its Two Micron All-Sky Survey (2MASS,
Skrutskie et al. \cite{Skrutskie}) $K_{\rm S}$-band magnitude. The 
signal-to-noise ratio (S/N) of the final spectrum is $\sim 60$. The 
heliocentric velocity correction was carried out and the spectrum was 
dereddened with the extinction value listed in Table\,\ref{Tab:param} 
applying the interstellar reddening law of Howarth (\cite{Howarth}).

A high-resolution spectrum ($R\simeq 50000$) in the $K$-band 
($2.276-2.326\,\mu$m) was obtained on 2009 December 2 using the CRyogenic 
high-resolution InfraRed Echelle Spectrograph (CRIRES; Kaeufel et al. 
\cite{Kaeufl}). The observations were performed with a slit-width of 
0.4\,arcsec, and a standard nodding on-slit strategy was applied to remove sky- 
and detector glow. For telluric corrections, a standard star was 
observed close in airmass and directly after the science object. The data 
reduction was performed with the CRIRES pipeline. The final spectrum was 
corrected for heliocentric velocity, and it has an S/N of $\sim 120$.

\section{Results}\label{results}

\begin{figure}[t!]
\resizebox{\hsize}{!}{\includegraphics{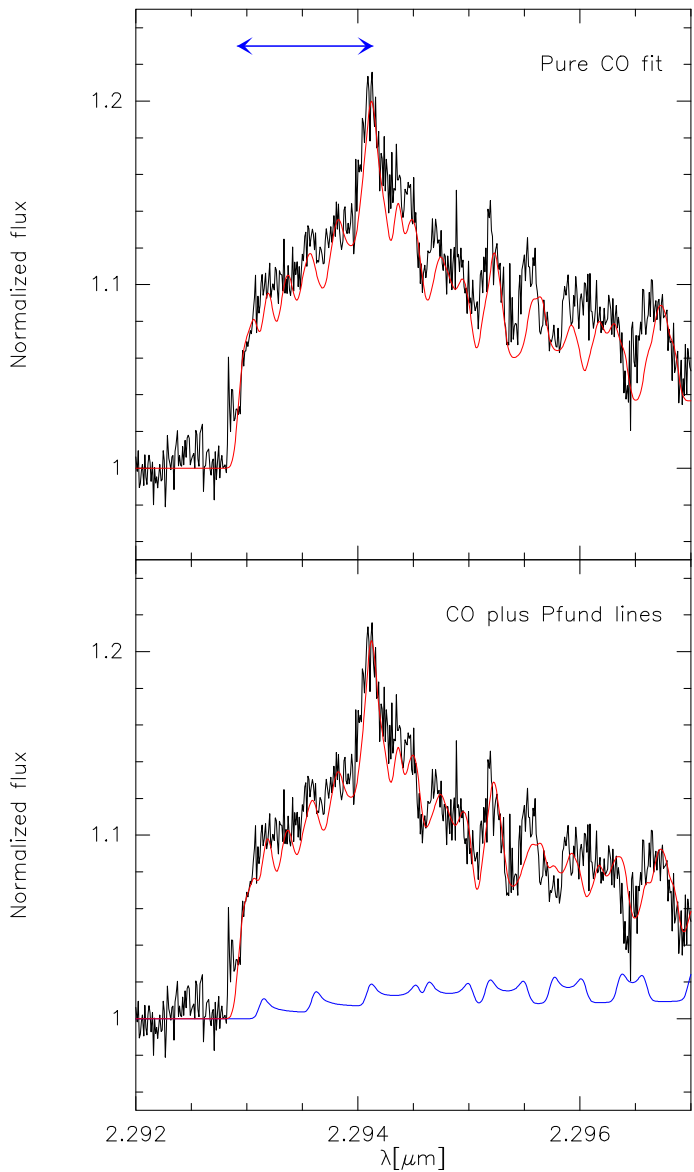}}
\caption{Fit (red/gray) to the observed, normalized CRIRES spectrum of 
GG\,Car (black) considering the emission to consist of either pure CO band 
(top panel) or CO band plus Pfund line emission (bottom panel).}
\label{Fig:CRIRES}
\end{figure}

In addition to the emission from both the \element[][12]{CO} and 
\element[][13]{CO} bands, the SINFONI spectrum (Fig.\,\ref{Fig:SINFONI}) 
clearly shows emission 
lines from the hydrogen Pfund series. To model their contribution, we first
adopted a Gaussian-like profile, indicative of an origin within a wind or 
shell. However, no reasonable fit of their profile shape could be achieved. 
%Instead, the observed Pfund lines appear slightly double-peaked. Other clearly 

Double-peaked lines have been identified in the optical spectra of GG\,Car, 
e.g., forbidden emission lines of [\ion{O}{i}] and [\ion{Ca}{ii}] (Muratore et 
al. \cite{Flor2012}) and permitted emission lines of \ion{Fe}{ii} (Marchiano 
et al. \cite{Paula}). Furthermore, emission lines from the hydrogen Balmer and 
Paschen series appear to be double-peaked, typically with the red peak more 
intense than the blue one (Marchiano et al. \cite{Paula}). Whether the Pfund
lines show the same behavior is hard to tell because of the poor quality of the
SINFONI spectrum (see noise level shortward of $2.29\,\mu$m).

Pfund emission is usually formed within a narrow region. Considering that the 
line profiles could be double-peaked, the emission might arise from a (narrow) 
ring within the rotating disk. The spectral resolution of SINFONI, $\varv_{\rm 
res} \simeq 60$\,km\,s$^{-1}$, smears out the possibly double-peaked profile 
shape so that we cannot measure the proper rotation velocity from the peak 
separation. The rotation speed, projected to the line of sight, was therefore 
obtained from fitting the line widths with a rotation profile convolved to the 
proper spectral resolution. The optimum fit was achieved for $\varv_{\rm rot, 
Pf} \sin i = (94 \pm 2)$\,km\,s$^{-1}$.

The Pfund series reaches its edge (i.e., the Pfund discontinuity) blueward of 
the CO band head. Furthermore, the higher the transition, the shorter the 
wavelength separation between individual lines so that the lines start to 
blend. The broad profiles of the Pfund lines in our spectra together with the 
low spectral resolution result in the formation of a quasi-continuum of the 
higher transitions of the Pfund series, i.e., in the wavelength region $\lambda 
\la 2.325\,\mu$m and extending blueward of the CO band head (see, e.g., Kraus 
et al. \cite{CO}). However, the SINFONI spectrum does not show any indication 
for such a quasi-continuum on the short wavelength edge of the 
CO spectrum, so the number of Pfund lines in the spectrum must be limited. 
This agrees with the fact that population of the higher levels in 
especially high-density media is usually prevented by pressure ionization 
effects. The maximum Pfund transition cannot be determined purely based on the 
SINFONI data, but the highest Pfund line should occur redward of the CO band 
edge.

The CRIRES spectrum (Fig.\,\ref{Fig:CRIRES}) shows the fully resolved structure 
of the first \element[][12]{CO} band head emission. While fitting this short 
portion 
of the emission spectrum is not sensitive to the CO temperature and column
density, it delivers detailed knowledge of the CO kinematics. The  
blue shoulder and red peak are typical characteristics for emission in a 
rotating ring or disk (e.g., Carr et al. \cite{carr93}; Carr \cite{carr95}; 
Najita et al. \cite{najita}; Kraus et al. \cite{CO}), and their separation 
(as shown by the arrow in the top panel of Fig.\,\ref{Fig:CRIRES}) directly 
determines the disk rotational velocity, projected to the line of sight, in the 
CO-forming region. 

The observed band head characteristics could also be interpreted with an 
equatorial outflow in the form of a gaseous ring expanding with constant
velocity (e.g., Kraus et al. \cite{KBA2010}), as the resulting double-peaked 
profile is indistinguishable from the rotating one. Nevertheless, we consider 
here the scenario of a rotating disk as more reliable based on the studies of 
Pereyra et al. (\cite{Pereyra}).

From our spectrum we obtain
$\varv_{\rm rot, CO} \sin i = (80 \pm 1)$\,km\,s$^{-1}$. This value is lower
than that obtained from the width of the Pfund lines and hence 
perfectly agrees with the scenario of a Keplerian rotating disk in which
the CO-emitting region is located farther away than the Pfund-emitting region.
From the projected velocity of the CO gas we obtain the actual rotational 
velocity of $\varv_{\rm rot, CO} = (91.5\pm 7.5)$\,km\,s$^{-1}$, assuming that 
the disk is co-planar with the orbital orientation. 

For the model computations, we made use of the CO disk code developed by Kraus 
et al. (\cite{CO}), modified to account for the emission of \element[][13]{CO}
(Kraus \cite{Kraus09}; Oksala et al. in preparation). Former calculations
of CO emission from a disk revealed that it is sufficient to consider the
innermost ring of gas with a constant temperature, column density, and rotation
velocity (e.g., Kraus \cite{Kraus09}; Liermann et al. \cite{Liermann}; Cidale
et al. \cite{Lydia}).

Fitting the CRIRES spectrum with pure CO emission did not deliver a reasonable 
fit, because several features of the spectrum could not be satisfactorily
reproduced (see top panel of Fig.\,\ref{Fig:CRIRES}). Therefore,
emission from the rotationally broadened Pfund lines was included. The best
fit (bottom panel of Fig.\,\ref{Fig:CRIRES}) was obtained for a cut-off in the
Pfund series at the transition Pf\,63. The contribution of the Pfund emission
is shown in this figure as well.

\begin{table}
\caption{Parameters of the CO disk.}
\label{Tab:CO}
\centering
\begin{tabular}{ll}
%\begin{tabular}{rrcrrr}
\hline
\hline
$T_{\rm CO}$ [K]                      & $3\,200\pm 300$ \\
$N_{\rm CO}$ [cm$^{-2}$]              & $(5\pm 3)\times 10^{21}$ \\
$i$ [$\degr$]                         & $63\pm 9$ \\
\element[][12]{C}/\element[][13]{C}   & $15\pm 5$ \\
$\varv_{\rm rot, CO}$ [km\,s$^{-1}$]  & $91.5\pm 7.5$ \\
$\varv_{\rm rot, Pf}$ [km\,s$^{-1}$]  & $108\pm 11$ \\
$A_{\rm CO}$ [AU$^{2}$]               & $1.33\pm 0.73$\\
%     $[K]$   & [cm$^{-2}$] & [$\degr$] & [km\,s$^{-1}$]  &  &  [AU$^{2}$]  \\
%\hline
%$3\,200\pm 300$ & $(5\pm 3)\times 10^{21}$   & $63\pm 9$  & $91.5\pm 7.5$ & $15\pm 5$ & $1.33\pm 0.73$ \\
%$\pm 300$ & $\pm 3\times 10^{21}$ & $\pm 9$ & $\pm 7.5$ & $\pm 5$ & $\pm 0.73$\\
\hline
\end{tabular}
\tablefoot{The inclination angle is taken from Marchiano et al. (\cite{Paula}).}
\end{table}

%\begin{table}
%\caption{Parameters for the best-fitting CO model. The second row
%gives the errors for each parameter.}
%\label{Tab:CO}
%\centering
%\begin{tabular}{cccccc}
%%\begin{tabular}{rrcrrr}
%\hline
%\hline
%$T_{\rm CO}$ & $N_{\rm CO}$ & $i$  & $\varv_{\rm rot}$ & \element[][12][C]/$^{13}$C & $A_{\rm CO}$\\
%     $[K]$   & [cm$^{-2}$] & [$\degr$] & [km\,s$^{-1}$]  &  &  [AU$^{2}$]  \\
%\hline
%3\,200      & $5\times 10^{21}$   & 63      & 91.5        & 15       & 1.33 \\
%$\pm 300$ & $\pm 3\times 10^{21}$ & $\pm 9$ & $\pm 7.5$ & $\pm 5$ & $\pm 0.73$\\
%\hline
%\end{tabular}
%\end{table}

With this cut-off in the Pfund series and fixing the kinematics in both
the CO and Pfund line emitting regions, we were finally able to obtain
reasonable fits to the total SINFONI spectrum (Fig.\,\ref{Fig:SINFONI}).
The complete set of derived CO disk parameters is
listed in Table\,\ref{Tab:CO}. The last row gives the area of
the emitting CO ring considering the disk inclination and the distance of
$d=5\pm 1$\,kpc towards GG\,Car as obtained by Marchiano et al. (\cite{Paula}).

The enrichment in \element[][13]{C} obtained from the strength of the 
\element[][13]{CO} band 
emission of GG\,Car is lower than postulated by Kraus (\cite{Kraus09}), who 
suggested \element[][12]{C}/\element[][13]{C} $< 10$. This disagreement can be 
explained by the much lower resolution of the AMBER spectrum in which the 
strongly rotationally broadened Pfund lines, which are even in the SINFONI 
spectrum fully resolved only at the long-wavelength end, are completely smeared 
out. Still, their presence pushes the \element[][13]{CO} band heads up, 
falsely suggesting a stronger enrichment than the actual value of 
\element[][12]{C}/\element[][13]{C} $= 15\pm 5$ found with our detailed 
modeling. 
  
The derived CO parameters (i.e., temperature and column density) are similar to
those found for other B[e]SGs such as LHA\,120-S\,12 and LHA\,120-S\,73 in the
Large Magellanic Cloud (Liermann et al. \cite{Liermann}), LHA\,115-S\,65 in the
Small Magellanic Cloud (Oksala et al. \cite{Mary}), and CPD-52\,9243 (Cidale et
al. \cite{Lydia}) and MWC\,349A (Kraus et al. \cite{CO}) in the Galaxy. Only
one Galactic object, HD\,327083, seems to be extreme in the sense that its
disk, studied by Wheelwright et al. (\cite{Wheelwright}), has the coolest
($T_{\rm CO} = 1721\pm 540$\, K) and densest ($N_{\rm CO} = (0.4\pm 1.8)\times
10^{24}$\,cm$^{-2}$) CO gas. However, in the case of HD\,327083 only the first
band head was studied, which is a poor tracer for density and
temperature. These parameters can be determined with a high precision only from
a complete CO band covering three or more band heads.

\section{Discussion}\label{disc}

\subsection{Circumbinary disk of GG\,Car}

The binary parameters derived by Marchiano et al. (\cite{Paula}) resulted in a 
size of the semi-major axis of the system (corrected for inclination) of $A = 
0.65\pm 0.04$\,AU and individual stellar masses of $M_{\rm prim} \simeq 26\pm 
4\,M_{\odot}$ and $M_{\rm sec} \simeq 12\pm 2\,M_{\odot}$. Interpreting the 
velocity of the CO gas with rotation in a Keplerian disk, the distance of the 
CO emitting ring would be $R_{\rm CO} = 2.88\pm 0.89$\,AU when orbiting the 
primary, and $R_{\rm CO} = 1.35\pm 0.45$\,AU when orbiting the secondary star. 
In both cases, this distance is larger than the binary separation. The CO must 
therefore be located in a circumbinary disk orbiting a central mass of 
$M_{\rm binary} \simeq 38\pm 6\,M_{\odot}$. 

\begin{figure}[t!]
\resizebox{\hsize}{!}{\includegraphics{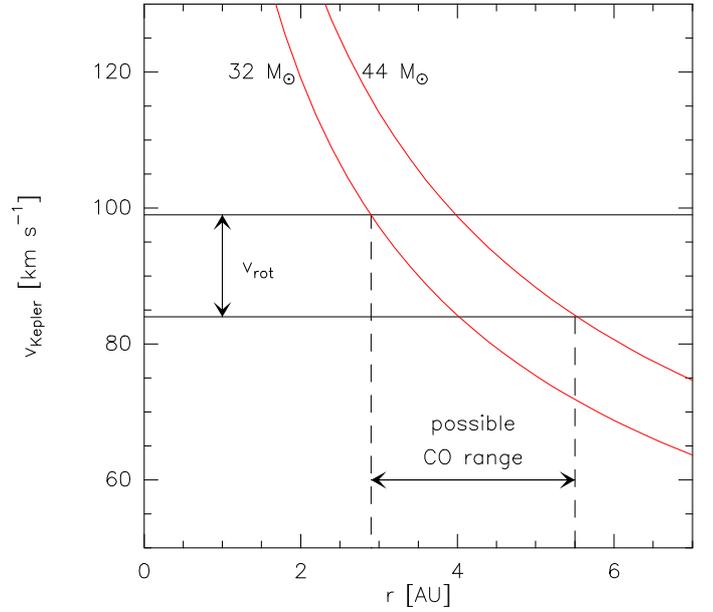}}
\caption{Keplerian velocity distribution considering the total
minimum and maximum binary mass. Horizontal lines represent the range in
rotational velocity of the CO gas. The location range of the CO ring is
marked (dashed lines).}
\label{Fig:vkep}
\end{figure}

Fig.\,\ref{Fig:vkep} displays the Keplerian velocity as a 
function of distance for the minimum and maximum binary masses. The horizontal 
lines show the range in rotation velocity obtained from modeling the CRIRES 
data, and the intersections deliver the range of possible distances of the 
circumbinary CO-emitting ring from the center of gravity. From the size of the 
emitting CO area (Table\,\ref{Tab:CO}) the outer edge, $R_{\rm out}$, of the CO 
emitting ring can be calculated as a function of its inner edge, $R_{\rm in}$. 
Considering the error in the CO area, we computed the two limiting scenarios
and obtained sizes for the CO ring that
range from $R_{\rm out} = 2.97\pm 0.04$\,AU for the minimum inner edge of 
$R_{\rm in, min} = 2.90$\,AU, and $R_{\rm out} = 5.54\pm 0.02$\,AU for the 
maximum inner edge of $R_{\rm in, max} = 5.50$\,AU, indicating that the 
emitting CO gas is confined in a narrow ring. Considering even the 
largest distance, the CO-emitting ring falls well inside the value of 230 
stellar radii ($\sim 34$\,AU for a primary stellar radius of 
$32\,R_{\odot}$) suggested by Marchiano et al. 
(\cite{Paula}) as the maximum possible distance of the CO-emitting region
based on temperature arguments. 

The B[e]SG star HD\,327083 was also recently found to have a circumbinary 
rotating disk displaying strong CO band emission (Wheelwright et al. 
\cite{Wheelwright}). 
%Likewise, for GG\,Car we do not see any indication for 
%deviation from a circular orbit. Furthermore, 
In both this and the GG\,Car system, the minimum 
inner edge of the CO ring is located at a similar distance of $\sim 3$\,AU. 
%The elliptic orbit of the CO gas in 
%HD\,327083 might be related to the fact that the binary separation and orbital 
%period in HD\,327083 ($A\simeq 1.7$\,AU; $P\sim 6$\,months) are substantially 
%larger than those found for GG\,Car ($A\simeq 0.65$\,AU; $P\sim 31$\,days) so 
%that the orbit of the circumbinary material could be influenced.
%
Other interesting objects in which a close (in projection) binary component was 
discovered by means of interferometry are the A[e] supergiant HD\,62623 
(Millour et al. \cite{Millour11}, although this star was recently re-classified 
as a B8-9 Ib-II star by Borges Fernandes et al. in prep.) and the B[e]SG 
candidates HD\,87643 (Millour et al. \cite{Millour09}) and V921\,Scorpii (=
CD-42\,11721) (Kraus et al. \cite{Stefan}). Only the first two have been found 
to display CO band emission, and detailed modeling will determine the
kinematics of the CO gas and possibly help to distinguish 
between the scenarios of circumstellar or circumbinary origin (Muratore et al. 
in preparation).

\subsection{Evolutionary stage of GG\,Car}

The origin of the B[e] phenomenon in massive evolved stars like B[e]SGs is 
still not fully resolved. A disk or ring of high-density material containing 
gas and dust is usually seen around these objects, but the disk formation 
mechanism is not fully understood and the disk kinematics have not been 
sufficiently studied yet. Because post-main sequence massive stars possess 
strong line-driven stellar winds during their main-sequence evolution, these 
disks cannot be remnants from their pre-main sequence stage, but likely formed 
from highly non-spherically symmetric stellar mass-loss. A possible scenario 
discussed in the literature to explain the formation of outflowing high-density 
disks in single-star evolution is the bistability mechanism induced by rotation 
(Lamers \& Pauldrach \cite{LamersPauldrach}; Pelupessy et al. 
\cite{Pelupessy}). However, the density of these disks increases enormously 
when the slow-wind solution (Cur\'{e} \cite{Cure}) at high rotation rates 
($\Omega > 0.6$) is considered (Cur\'{e} et al. \cite{Cureetal}). In both, the 
rapid stellar rotation is an important prerequisite, but it is even more 
important to know if a massive main-sequence star could remain 
close to the limit of critical rotation for a substantial fraction 
of its lifetime (Langer \cite{Langer}).
Rotation velocities at a substantial fraction of their critical 
values have been found so far for two B[e]SGs LHA\,115-S\,23 (Kraus et al. 
\cite{KBKA}) and LHA\,115-S\,65 (Zickgraf \cite{Zickgraf}; Kraus et al. 
\cite{KBA2010}) in the Small Magellanic Cloud. But whether all B[e]SGs are 
indeed rapid rotators is not known.

\begin{figure}[t!]
\resizebox{\hsize}{!}{\includegraphics{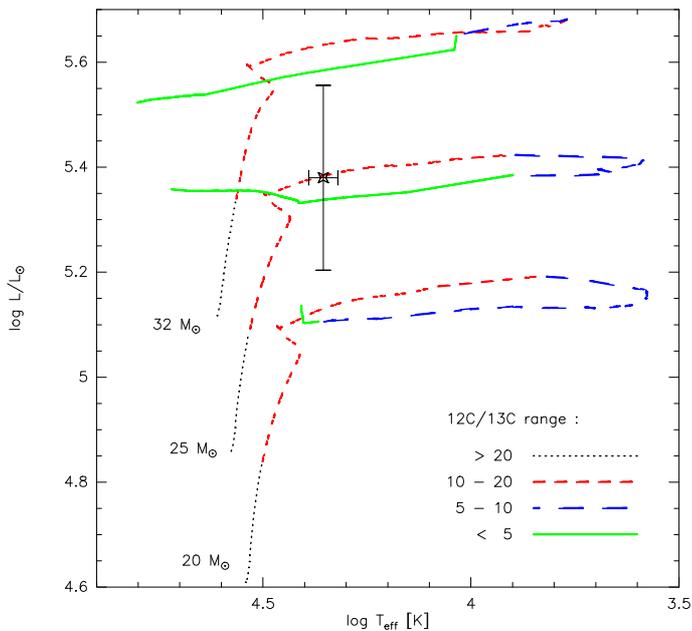}}
\caption{HR diagram showing the position of the primary component of GG\,Car
with respect to new evolutionary tracks of Ekstr\"{o}m et al. (\cite{Ekstrom})
for stars at solar metallicity rotating initially with $\varv_{\rm ini} /
\varv_{\rm crit} = 0.4$. The enrichment in \element[][13]{C} over the course of 
stellar evolution is indicated.}
\label{Fig:HRD}
\end{figure}

Other scenarios consider that binary interaction up to full merging of massive 
binaries can cause the B[e] phenomenon in the supergiants (Langer \& 
Heger \cite{LangerHeger}; Podsiadlowski et al. \cite{Podsi}). However, the
binary merger scenario seems to hold so far for only one object of this class, 
i.e., the B[e]SG R4 in the Small Magellanic Cloud. 

The appearance of the B[e] phenomenon in the eccentric binary system GG\,Car 
might also be linked to binary interaction, at least to some extent, and in the 
following we briefly discuss the evolutionary stage of the primary component of 
GG\,Car and its possible evolutionary history.

Figure \ref{Fig:HRD} shows the location of the primary component of GG\,Car in 
the HR diagram based on its effective temperature and luminosity 
(Table\,\ref{Tab:param}). Also shown are new tracks from Ekstr\"{o}m et al. 
(\cite{Ekstrom}) for single-star evolution at solar metallicity and initial
rotation speeds of about 40\% of the critical velocity. The rotation speed
might not be fully appropriate for GG\,Car, for which no rotational velocity
could be obtained so far due to the lack in photospheric lines. But an initial 
rotation speed like this has been found as the peak of the velocity distribution 
for young B stars (Huang et al. \cite{Huang}), and thus can be regarded as 
representative for the general behavior of rotating stars. The different line 
styles used along the evolutionary tracks in Fig.\,\ref{Fig:HRD} mark regions 
with different values of the \element[][12]{C}/\element[][13]{C} surface 
abundance ratio. This ratio starts with an initial (interstellar) value of 
$\sim 90$ on the zero-age main sequence, and drops to a final value of $\la 5$ 
at late, post-red supergiant evolutionary phases, tracing a strong enrichment 
in \element[][13]{C} during the evolution of massive stars. The value of the 
carbon isotopic ratio found for the circumbinary material of GG\,Car is 
\element[][12]{C}/\element[][13]{C}$ = 15\pm 5$. This high value excludes a 
late (i.e., post-red supergiant) evolutionary phase, but excellently 
agrees with the primary having just evolved beyond the main sequence (see 
Fig.\,\ref{Fig:HRD}) considering that the primary has evolved unaffected by 
its companion.

The initial mass and current location of GG\,Car in the HR diagram
is very similar to another B[e]SG, the object LHA\,120-S\,12 in the Large 
Magellanic Cloud. Furthermore, the CO ring around this star was found to have 
very similar CO column density ($N_{\rm CO} = (2.5\pm 0.5)\times 
10^{21}$\,cm$^{-2}$) and temperature ($2800\pm 500$\,K), and its carbon 
isotopic ratio was determined to $20\pm 2$ (Liermann et al. \cite{Liermann}). 
Like LHA\,120-S\,12, GG\,Car is hence another example of a B[e]SG with only 
slight enhancement of \element[][13]{C} and consequently is in a pre-red 
supergiant phase, while other objects were found to show strong 
\element[][13]{C} enrichment, which classifies them as post-red supergiants or 
post-yellow hypergiants (see e.g., Liermann et al. \cite{Liermann}; Muratore et 
al. \cite{Flor2010}; Oksala et al. in preparation). 

\begin{figure}[t!]
\resizebox{\hsize}{!}{\includegraphics{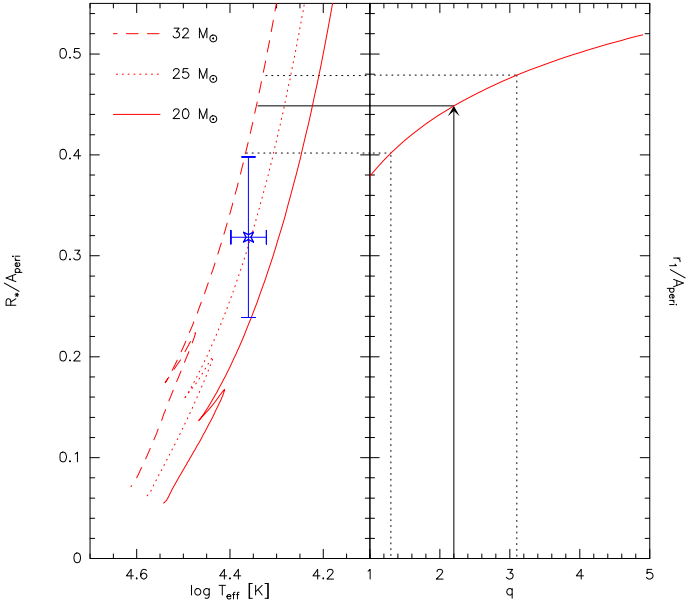}}
\caption{Left panel: Stellar radius, normalized to the minimum binary 
separation of the GG\,Car system, during the main and post-main sequence 
evolution of stars with different initial masses, plotted in terms of the 
effective temperature variation. 
Right panel: Relative effective Roche lobe radius for a range of mass ratios. 
The value (with errors) for GG\,Car is marked.}  
\label{Fig:roche}
\end{figure}

Although the carbon isotopic ratio is the strongest indicator, there are
additional arguments in favor of an evolutionary phase of GG\,Car just beyond 
the main sequence. Unlike the well-known LBVs HR\,Car and AG\,Car, no 
nebulosity was detected around GG\,Car (Thackeray \cite{Thackeray}). Therefore, 
it seems less evolved than its close-by neighbors. Furthermore, assuming that 
the three stars HR\,Car, AG\,Car and GG\,Car have similar ages (based on 
similar sky location and distance), it is clear 
that the more massive stars are more evolved. Comparing the ages of AG\,Car and 
HR\,Car with the evolutionary tracks of stars in the proper mass range for the 
primary of GG\,Car delivers an evolutionary stage just beyond the 
main sequence, in perfect agreement with its position in the HR diagram.

In summary, all evidence (\element[][13]{C} enrichment, age, missing LBV 
nebulosity) seems to speak in favor of a single-star evolutionary scenario for 
the primary component of GG\,Car, which is currently in an early post-main 
sequence evolutionary phase, according to the HR diagram.

\subsection{Origin of the circumbinary disk}

One question that remains to be discussed concerns the origin of the cool and 
dense circumbinary disk seen in both CO and dust emission. The transportation 
and accumulation of stellar material into circumbinary orbit might have 
happened in two different ways, either via non-conservative Roche lobe overflow
occurring earlier in the evolution of the primary, or the primary 
became a Be star during its late stages of main-sequence evolution. In 
the following we briefly describe and discuss both options.

Roche lobe overflow results when the stellar radius becomes larger than the 
size of its Roche lobe. This can happen at different stages of stellar 
evolution.
To compute the effective radius, $r_{\rm 1}$, of GG\,Car's Roche lobe we 
applied the equation
\begin{equation}
\frac{r_{1}}{A} = \frac{0.49 q^{2/3}}{0.6 q^{2/3} + \ln(1+q^{1/3})}
\label{radius}
\end{equation}
provided by Eggleton (\cite{Eggleton}). The parameter $q$ is the mass ratio
and $A$ is the binary separation. This equation is strictly valid for 
circular orbits only. In eccentric binaries, the mass transfer will take place
when the two stars are closest to each other, i.e., during periastron passage.
We can hence approximate an equivalent Roche lobe for eccentric binaries
by replacing the binary separation $A$ by the periastron distance $A_{\rm
peri} = A (1-e)$, where $e$ is the eccentricity. The behavior of $r_{1}/A_{\rm
peri}$ as a function of the mass ratio $q$ is shown in the right panel of
Fig.\,\ref{Fig:roche}. For the chosen range, the radius of the Roche lobe
increases with increasing $q$. The mass ratio (including its errors)
of GG\,Car and the resulting size of the effective Roche lobe is indicated.

To determine if the radius of GG\,Car's primary star could have filled its
Roche lobe during its past evolution, we computed the evolution of the stellar
radius during and beyond the main sequence for stars with different appropriate 
initial masses. Assuming that the periastron distance remained constant
during the star's evolution, we normalized the stellar radii to GG\,Car's 
current periastron distance and plot them versus the evolution of the 
stellar effective temperature in the left panel of Fig.\,\ref{Fig:roche}.
The location of GG\,Car's primary is included in the plot, as well as the
range in Roche lobe radii for the current mass ratio. Obviously, as long 
as neither the binary separation nor the mass ratio has changed during 
earlier evolution, the primary of GG\,Car has not filled its Roche lobe
yet. If, however, we postulate that the circumbinary material originates
from non-conservative Roche lobe overflow in the past, then the original
mass ratio must have been higher than the current value. A higher mass
ratio, on the other hand, increases the effective radius of the Roche lobe.
Hence, for mass transfer to happen, the original periastron distance must 
have been (much) smaller than it is today, and the mass transfer must have
resulted in widening the system.

While this scenario cannot be excluded (see, e.g., Sepinsky et al. 
\cite{Sepinsky07, Sepinsky09}), non-conservative mass transfer usually leads 
to a shrinking orbit. In addition, standard theory of binary evolution assumes 
that the originally eccentric orbit circularizes immediately after the
onset of mass transfer (e.g., Hurley et al. \cite{Hurley}). This assumption 
seems to be justified because only few eccentric evolved binary systems 
are observed (e.g., de Mink et al. \cite{Selma}).
In this context, the eccentric orbit seen for GG\,Car might speak in 
favor of an unevolved binary system in which no mass transfer has occurred 
yet, which would agree with the primary radius being 
smaller than the effective Roche lobe size during periastron passage (see 
Fig.\,\ref{Fig:roche}).

However, Roche lobe overflow cannot be entirely excluded, because
Fig.\,\ref{Fig:roche} was computed for the mean value of the periastron 
distance. Taking into account its error, which is based on the error in 
inclination angle and eccentricity, and considering the extreme case of
minimum mass ratio and maximum stellar radius, the effective radius of
the Roche lobe drops below the stellar radius. Therefore, Roche lobe overflow 
could be possible, but would be restricted to an extreme corner in the overall 
parameter space.
 
In the second possible scenario, the primary of GG\,Car could have evolved from 
a Be star progenitor, particularly if its mass is closer to the lower 
boundary of the primary's mass range. This value would agree better
with those of massive Be stars found close to the end of and slightly beyond 
the main sequence. Be stars show Balmer emission 
that is supposedly arising from a geometrically thin, ionized gaseous disk in 
Keplerian rotation (e.g., Porter \& Rivinius \cite{PorterRivi}). The disk is 
thought to be formed by mass loss through an equatorial, viscous decretion disk 
(Lee et al. \cite{Lee}; Porter \cite{John}; Krti\v{c}ka et al. \cite{Jiri}) 
triggered by the close to critical rotation of the star (e.g., Townsend et al.
\cite{Townsend}). Typical sizes range from a few up to a few hundred stellar 
radii.

Assuming that the plane of the disk around the primary is coplanar with the 
binary orbit, this disk can cause mass loss into 
circumbinary orbits in two ways. The binary separation of GG\,Car during 
periastron is $A_{\rm peri} = 0.468\pm 0.048$\,AU, which corresponds to 
roughly 5-15 times the stellar radius during the main sequence evolution
(see Fig.\,\ref{Fig:roche}). Therefore, during periastron passage even a
small disk will extend beyond the Roche radius and disk mass could stream
to the companion through the Lagrangian point and/or leave the Roche surface
to accumulate in circumbinary orbits. On the other hand, the secondary will 
have to pass through the disk as well, and disk material approaching the star
from behind will be accelerated and hence kicked out of the system.  

It is hard to tell which of the two suggested scenarios is real. But given 
the need for an extreme parameter combination for Roche lobe overflow to work, 
in combination with the eccentric orbit of GG\,Car that points to an unevolved 
binary system, the Be star scenario might be slighlty more favorable.
Nevertheless, only proper
theoretical modeling of both can give a more conclusive answer to the question
of the evolutionary history of GG\,Car. In addition, a  proper study of the 
kinematics of the circumstellar/circumbinary material needs to be performed 
based on many individual tracers of different regions. The easiest indicators 
for kinematics are certainly the forbidden emission lines.
Among the observed lines are [\ion{O}{i}] and [\ion{Ca}{ii}], which are
the most reliable tracers for the high-density disk regions (Aret et al.
\cite{Aret}). The profiles of the [\ion{O}{i}] lines, which seem to consist of 
two distinct double-peaked profiles (Muratore et al. \cite{Flor2012}), already 
show that the distribution and kinematics of the material
in the GG\,Car system is highly complex. High-quality observations, especially
over the full orbital period, are of vital importance to extract a more
complete picture of this fascinating object.

\section{Conclusions}\label{summary}

We presented our analysis of the CO-band emission of the eccentric binary 
system GG\,Car. CO bands are detected in both medium- and high-resolution 
$K$-band spectra obtained with the SINFONI and CRIRES spectrographs. The 
SINFONI spectrum extends over the full CO-band structure, which gives insight 
into the density ($N_{\rm CO} = (5\pm 3)\times 10^{21}$\,cm$^{-2}$) and 
temperature ($T_{\rm CO}=3200\pm 500$\,K) of the emitting region, which
we find to be confined in a narrow circumbinary ring. In addition, the SINFONI 
spectrum covers the band heads from the isotopic molecule \element[][13]{CO}, 
allowing us to determine the evolutionary phase and hence the age of the star 
at the time of the material ejection. Our modeling derives a value 
\element[][12]{C}/\element[][13]{C} $= 15\pm 5$, confirming a slightly evolved 
(supergiant) nature of GG\,Car's primary component. The CRIRES spectrum 
contains the spectrally resolved structure of the first 
\element[][12]{CO} band head. From its shape we find that the 
motion of the CO gas is consistent with Keplerian rotation 
with a velocity, projected to the line of sight, of $80\pm 
1$\,km\,s$^{-1}$. 

Based on these results, we discussed the origin of the circumbinary disk in
this eccentric binary system. In particular, two scenarios were taken into 
consideration. The first one considered the possibility that the primary 
component underwent non-conservative Roche lobe overflow during its past 
evolution, while the second considered that the mass transport and
accumulation on circumbinary orbits could have happened during Be phases of 
the primary component. We cannot firmly prove or disprove the validity
of either scenario. However, the conditions for non-conservative
Roche lobe overflow are only met in an extreme corner of the parameter 
space. Therefore, slight preference is given to the suggestion that during 
late phases of its main-sequence evolution, GG\,Car's primary could have been 
a classical Be star.

%%%%%%%%%%%%%%%%%%%%%%%%%%%%%%%%%%%%%%%%%%%%%%%%%%%%%%%%%%%%%%%%%%%%%%%%%

\begin{acknowledgements}
We thank the anonymous referee for the valuable comments on the manuscript.
This research made use of the NASA Astrophysics Data System (ADS). M.K., 
M.E.O., D.H.N., and A.A. acknowledge financial support from GA\v{C}R under 
grant number P209/11/1198. The Astronomical Institute Ond\v{r}ejov is supported 
by the project RVO:67985815. Financial support for International Cooperation of 
the Czech Republic (M\v{S}MT, 7AMB12AR021) and Argentina (Mincyt-Meys, 
ARC/11/10) is acknowledged. 
M.F.M. is a research fellow of the Universidad Nacional de La Plata, Argentina.
M.B.F. acknowledges Conselho Nacional de Desenvolvimento
Cientifico e Tecnol\'{o}gico (CNPq-Brazil) for a post-doctoral grant.
A.A. acknowledges financial support from the research project SF0060030s08
of the Estonian Ministry of Education and Research.
L.C. acknowledges financial support from the Agencia de Promoci\'{o}n 
Cient\'{i}fica y Tecnol\'{o}gica (BID 1728 OC/AR PICT 885), PIP 0300 CONICET,
and the Programa de Incentivos G11/109 of the Universidad Nacional de La
Plata, Argentina.
\end{acknowledgements}

%%%%%%%%%%%%%%%%%%%%%%%%%%%%%%%%%%%%%%%%%%%%%%%%%%%%%%%%%%%%%%%%%%%%%%%%%

%\Online

\end{document}